\pdfoutput=1

\documentclass[sigconf]{acmart}

\usepackage[capitalise]{cleveref}
\usepackage{dirtytalk}
\usepackage[inline]{enumitem}
\usepackage{timing-diagrams}

\newcommand{\os}{\textsc{OS}}

\newenvironment{boxed_new}
    {\begin{center}
    \begin{tabular}{|p{0.45\textwidth}|}
    \hline\\
    }
    {
    \\\\\hline
    \end{tabular}
    \end{center}
    }

\acmConference[MobileSoft'22]{The 9th International Conference on Mobile Software Engineering and Systems}{May 22–-23, 2022}{Pittsburgh, PA, USA}

\begin{document}

\title{Quantifying Daily Evolution of Mobile Software Based on Memory Allocator Churn}

\author[1]{Gunnar Kudrjavets}
\orcid{0000-0003-3730-4692}
\affiliation{
   \institution{University of Groningen}
   \city{Groningen}
   \postcode{9712 CP}
   \country{Netherlands}}
\email{g.kudrjavets@rug.nl}

\author[2]{Jeff Thomas}
\affiliation{
    \institution{Meta Platforms, Inc.}
    \streetaddress{1 Hacker Way}
    \city{Menlo Park}
    \state{CA}
    \country{USA}
    \postcode{94025}}
\email{jeffdthomas@fb.com}

\author[3]{Aditya Kumar}
\orcid{0000-0001-6312-2898}
\affiliation{
    \institution{Snap, Inc.}
    \streetaddress{2772 Donald Douglas Loop N}
    \city{Santa Monica}
    \state{CA}
    \country{USA}
    \postcode{90405}}
\email{adityak@snap.com}

\author[4]{Nachiappan Nagappan}
\orcid{0000-0003-1358-4124}
\affiliation{
    \institution{Meta Platforms, Inc.}
    \streetaddress{1 Hacker Way}
    \city{Menlo Park}
    \state{CA}
    \country{USA}
    \postcode{94025}}
\email{nnachi@fb.com}

\author[5]{Ayushi Rastogi}
\orcid{0000-0002-0939-6887}
\affiliation{
   \institution{University of Groningen}
   \city{Groningen}
   \postcode{9712 CP}
   \country{Netherlands}}
\email{a.rastogi@rug.nl}

\begin{abstract}
The pace and volume of code churn necessary to evolve modern software systems
present challenges for analyzing the performance impact of any set of code changes.
Traditional methods used in performance analysis rely on extensive data
collection and profiling, which often takes days.
For large organizations utilizing Continuous Integration (\textsc{CI}) and
Continuous Deployment (\textsc{CD}), these traditional techniques
often fail to provide timely and actionable data.
A different impact analysis method that allows for more efficient detection
of performance regressions is needed.
We propose the utilization of user mode memory allocator churn as a \emph{novel approach to performance engineering}.
User mode allocator churn acts as a proxy metric to evaluate the relative change
in the cost of specific tasks.
We prototyped the memory allocation churn methodology
while engaged in performance engineering for an i{OS} version of application X.
We find  that calculating and analyzing memory allocator churn
\begin{enumerate*}[label=(\alph*),before=\unskip{ }, itemjoin={{, }}, itemjoin*={{, and }}]
    \item results in deterministic measurements
    \item is efficient for determining the presence of both individual performance
regressions and general performance-related trends
    \item is a suitable alternative to measuring the task completion time.
\end{enumerate*}
\end{abstract}



\maketitle

\section{Introduction}

Software evolution is a well-researched topic~\cite{tripathy_2014}.
The commonly agreed-upon definition of software evolution is a process
of continual change from a lesser or worse state to a higher or
better state~\cite{arthur_1988}.
The \emph{how} view of software evolution focuses on practical
aspects \say{that provide \emph{means} to direct, implement
and control software evolution}\mbox{~\cite[p. 4]{mens_2008}}.
We apply principles of software evolution
to tackle the problem of
understanding how the performance of key user scenarios in mobile applications
changes daily.

Current industry practices for popular mobile applications, such as {F}acebook or {I}nstagram, use \textsc{CI}/\textsc{CD}~\cite{kellner_mobilescale_2018,rossi_continuous_2016}.
\emph{Performance regressions are an evolution of software in an undesired direction}.
Users of mobile applications are sensitive to performance regressions.
Poor performance increases the likelihood of application
abandonment~\cite{zuniga_2019}.
Optimizing performance characteristics is the key
to user satisfaction and prolonged engagement with the application~\cite{hort_2021}.
Given the frequency of releases in a \textsc{CI}/\textsc{CD} environment,
engineers need to have the means
to detect, diagnose, and fix performance regressions within a \emph{limited time frame}.
Moreover, traditional techniques used in performance analysis rely on extensive
amounts of data and profiling~\cite{gregg_systems_2020,jain_art_1991}.
The data collection and profiling can take days,
rendering the initial diagnosis obsolete.

To analyze the performance of an application more efficiently,
we propose a simplified heuristic of using \emph{memory allocator churn}.
Churn is the number of calls to the memory allocator and the size of objects
allocated and released in the heap of the current process during a specific period.
We use the churn as a proxy metric to
\begin{enumerate*}[label=(\alph*),before=\unskip{ }, itemjoin={{, }}, itemjoin*={{, and }}]
    \item determine the presence of performance regressions
    \item evaluate their severity
    \item rank the order of performance investigations.
\end{enumerate*}
This paper is based on our experience prototyping the usage of allocator
churn for six months while working on performance engineering
for an i{OS} version of application X.

\section{Background and motivation}
\label{sec:background}

\subsection{Contemporary software development}

Modern software development methodologies such as \textsc{CI}/\textsc{CD} strive towards frequent release of updates.
That approach is different from the waterfall model, which has been dominant for decades~\cite{royce_1987,benington_1983}.
A final version of software within the waterfall model is released only after a long development cycle followed by a thorough testing period.
However, popular mobile and web applications such as {F}acebook, {I}nstagram, and {T}witter use \textsc{CI}/\textsc{CD}~\cite{kellner_mobilescale_2018,rossi_continuous_2016}.

Each application has a core codebase and several dependencies.
The development of all these components can involve hundreds to thousands of engineers.
Each engineer may contribute code to their respective codebases daily.
Any code change can modify the application's behavior, dependency graph, or performance .

Every application has a set of critical scenarios and phases of execution
which can influence its popularity and usage.
For example, boot time is a key performance metric for the
operating system (\os),
responsiveness during composing and publishing a tweet is one of the critical
scenarios for {T}witter,
and application start time is a crucial metric for {F}acebook~\cite{verma_optimizing_2015}.
Research shows that \emph{responsiveness}, or time the application takes
to react to a user's command is
critical to the application's reputation and overall success~\cite{yang_2013}.

\subsection{Software evolution and performance}

The waterfall model
dominant in the past meant that engineers had time to understand
the current state of the product and conduct a performance analysis on it.
Given the increased focus on the single trunk development model,
the ability to control the inclusion of code changes in the specific
daily builds of the application diminishes~\cite{winters_2020}.
With \textsc{CI}/\textsc{CD},
\begin{enumerate*}[label=(\alph*),before=\unskip{ }, itemjoin={{, }}, itemjoin*={{, and }}]
    \item the amount of daily code churn is high
    \item application release cadence is frequent
    \item the time frame to fix the performance-related issues has decreased.
\end{enumerate*}
Our research question becomes:

\begin{boxed_new}
	\textbf{RQ1}: What data and metrics can we use to understand the
	software's evolution since the last daily build~\cite{mccarthy_dynamics_2006}?
\end{boxed_new}

Insights from those metrics can then be used to isolate the code changes
responsible for performance regressions.
Understanding the root cause of regressions enables us to make required
modifications to software
to optimize its behavior in the target environment and for the critical scenarios.

\section{Data available for analysis}
\label{sec:solutions}

\subsection{Code changes, commit descriptions, and work items}

Modern source control systems such as {G}it and {M}ercurial use the
distributed model.
Engineers commit and test the code changes locally.
When new code is ready for integration
into the main development branch, either the code collaboration tool
(e.g., {C}ode{F}low, {C}ritique, {G}errit, or {P}habricator)
merges them after
passing the code review, or changes are manually pushed to the target branch.
One solution to analyze the performance impact is to read through each code change,
commit description,
and resolved work item since the last daily build.
This approach is impractical for most engineers working on complex systems due to the sheer volume of data
processing.
Our empirical observations about commit descriptions indicate that they are
succinct and do not describe every single change in detail.
Moreover, reading through every code change in each component is not
time-effective.

\subsection{General performance characteristics}

A typical build validation process executes several automated test
cases on each candidate build daily.
Some of these test cases measure general performance-related characteristics
of the application under a set of common usage scenarios.
For example, active thread count, maximum memory usage, and
the number of modules loaded.
These metrics, however,
indicate only \emph{generally} the trend of an application's
behavior.
We can approximate whether an application is \say{doing more work} or utilizing more resources for performance engineering purposes, but we cannot pinpoint specifics.

\subsection{Specific observable performance characteristics}

In software performance engineering~\cite{gregg_systems_2020}, we can
distinguish between the following broad categories of metrics:

\begin{itemize}
	\item \textbf{\textsc{CPU}}.
	It is critical to know how long a particular operation takes for metrics that measure user experience (e.g., responsiveness).
	To measure the duration of an operation, it is common to use \emph{elapsed time} (\emph{wall-clock time}).
	A typical \os\ (not real-time) scheduler operates on a per-thread basis and
	makes decisions about quantum (allowance of \textsc{CPU} time) assignments
	based on the behavior of \emph{all} executing \os\ threads.
	This scheduling approach makes conducting \emph{deterministic measurements}
	time-consuming (due to several iterations required) and in most cases requires a variety of hardware and
	in-depth profiler data to interpret the results correctly.

	\item \textbf{\textsc{I/O}}.
	\textsc{I/O} refers to the interactions with the storage device
	such as a disk, inter-process communication, and network traffic.
	Slow \textsc{I/O} directly contributes to increased wall clock
time, impacting the user experience.
	Measuring the amount of \textsc{I/O} and speed is highly dependent on the environment
	(e.g., file system cache behavior on a particular \os),
	the hardware used, and application usage scenarios.

	\item \textbf{Memory}.
	Reducing memory consumption is a well-known optimization technique
	for mobile software~\cite{hort_2021}.
	Decreasing the amount of memory an application consumes
	reduces the application's launch time~\cite{lee_2021}.
	Metrics such as overall consumption, allocation rate,
	fragmentation, and page faults, are the basis of evaluation to
	determine the cost of scenarios.
	However, because of a variety of different classifications,
	even interpreting the exact meaning of a metric and impact to
	application's performance is an involved and costly process.

	\item \textbf{Battery or power}.
	Battery consumption is a metric relevant
	mainly to software running on mobile devices. Measuring power
	utilization correctly for intervals lasting tens or hundreds of milliseconds
	typically requires special hardware and access to facilities of the \os, which
	enable exposing that data.
	We find that this approach is not practical for daily performance analysis
	due to the investment of time required.
\end{itemize}

Gathering valid data about all these categories is a time-consuming process.
To cover a wide range of target environments, the performance testing needs to take
place on multiple versions of \os\ executing on different hardware versions.

A key observation, based on our industry experience when conducting
performance analysis
on a variety of products on different \os s is the following:

\begin{boxed_new}
	Most types of \say{work} done by any application
	require interaction with the memory manager.
    The efficiency of the allocator influences software performance~\cite{lever_2000}.
	\emph{Memory allocator churn} can be a deterministic
	way to measure the amount of work an application must do
	and is independent of \textsc{CPU} execution speed.
\end{boxed_new}

\section{Measuring memory allocator churn}
\label{sec:measuring}

\subsection{Overview}

Lower-level languages, such as {C}, {C}++, and earlier versions of {O}bjective-{C}
require explicit memory management.
Source code gives the reader details about memory allocation and release.
In higher-level programming languages such as {J}ava{S}cript, \textsc{PHP}, or {P}ython,
engineers are not required to manage memory explicitly.
An engineer may use data types like strings or hash tables
and never allocate or free memory associated with
them directly.
The performance costs for calling frameworks or utilizing
language constructs (e.g., blocks in {O}bjective-{C}) are not immediately visible based on
the code.
Therefore, it is hard for both the author and reviewer to understand the
performance impact of changes.

For this paper, we consider the standard \textsc{POSIX} memory
allocation functions (\texttt{free()}, \texttt{calloc()}, \texttt{malloc()}, and \texttt{realloc()}) as an interface to allocator~\cite{posix}.
The runtime for a particular language ends up eventually translating
various language constructs to use system libraries which consequently call
these functions.

\subsection{Metrics}

We define the critical parts of application execution as two marker
points in the execution timeline of a specific thread: $m_{k}$ and $m_{k+1}$.
For example, placing a starting marker at the beginning of the function
and the ending marker before the function returns.
Similarly, this approach is extendable to cover more involved user scenarios
like the start of rendering some content and when the rendering finishes.
Each marker can contain other child markers, and the markers can overlap.

Tasks like manipulating strings, loading modules, processing images,
calling other \textsc{API}s, and marshaling data
all involve allocating and deallocating memory.
We can quantify how expensive a particular critical phase is
by counting the amount of allocator churn during it.
To measure the allocator churn, we need to gather data about allocator usage.
Multiple ways to intercept calls to the allocator exist.
Using
callbacks~\cite[p. 977]{singh_mac_2016},
hooks~\cite{malloc_hooks},
preloading the shim~\cite{kobayashi_tips_2013}, and
debug memory libraries~\cite{dmalloc}
are all valid options.

We can quantify the allocator churn by looking at the
following values:

\begin{itemize}
	\item \emph{Total churn in bytes}.
	Allocating more memory
	will cause the allocator to execute more \textsc{CPU} instructions.
	For example, coalescing neighboring blocks, paging out some memory
	to the disk, and making syscalls.

	\item \emph{Total number of function calls to the allocator}.
	Calling more allocator functions means similarly execution of extra
	\textsc{CPU} instructions to validate the input,
	manage the stack frame, and
	acquire necessary synchronization primitives.
\end{itemize}

Anecdotally, we observe that both values are an indication of
the amount of work the caller's code will perform.
Putting the two observations above together lead us to propose calculating
the cost between two markers $m_{k}$ and $m_{k+1}$
on a specific thread $t_{j}$ as
\begin{equation}
churn\left(m_{k}, m_{k+1}\right) = \sum_{i=1}^{n} c\left(f_{i}, b_{i}\right)
\end{equation}

where $n$ is the number of memory allocation related functions
called between two markers,
$c$ is a function calculating the cost of a function $f_{i}$,
and $b_{i}$ is the number of bytes
the function $f_{i}$ operates on.

\subsection{Relative costs of allocator functions}

We define $c\left(f_{i}, b_{i}\right) = w\left(f\right)\cdot \log_{2} b_{i}$ where
$w$ is a weight for a function $f_{i}$ and $b_{i}$ is the number of bytes
the function $f_{i}$ operates on.
Defining the weight function $w$ depends on several variables:
\os\ and its version, types of compiler optimizations enabled, or
memory allocator used.
A simplified model assumes that allocating memory costs more
than freeing it, and reallocation is even more costlier than either allocation
or deallocation.

One possible initial definition is to use
$calloc() = 2$, $free() = 1$, $malloc() = 1$, and $realloc() =3$.

\subsection{Issues associated with valid measurements}

This section describes the conceptual issues associated with gathering valid measurements for the allocator churn.

\subsubsection*{Single thread versus multiple threads}

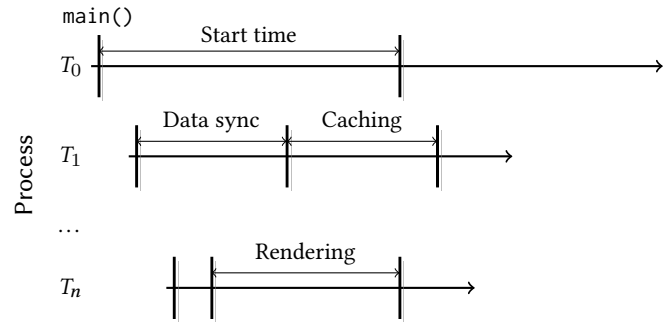
\begin{figure}[!htbp]
\centering
\begin{tikzpicture}
  \draw (-1,0.75)  node[rotate=90] (process) {\textsf{\large Process}};

  \tline{A}{2.2};
  \tcaption{A}{$T_{0}$};
  \tline{B}{1};
  \tcaption{B}{$T_{1}$};
  \tline{X}{0};
  \tcaption{X}{\ldots};
  \tline{C}{-0.75};
  \tcaption{C}{$T_{n}$};

  \ttimeline{A}{7.5};
  \tskip{B}{0.5};
  \ttimeline{B}{5};
  \tskip{C}{1};
  \ttimeline{C}{4};

  \ttick{A};
  \ttextU{A}{\texttt{main()}};
  \tskiptext{A}{4}{Start time};
  \ttick{A};

  \ttick{B}
  \tskiptext{B}{2}{Data sync};
  \ttick{B};
  \tskiptext{B}{2}{Caching};
  \ttick{B};

  \ttick{C}
  \tskip{C}{0.5};
  \ttick{C};
  \tskiptext{C}{2.5}{Rendering};
  \ttick{C};
\end{tikzpicture}
\caption{Process execution timeline with multiple threads $T_{i}$. Each thread has one or more critical phases that may overlap.}
\label{fig:timeline}
\end{figure}

An application can have multiple threads executing at the same time.
\Cref{fig:timeline} shows a sample snapshot of process execution.
Decisions made by the \os\ scheduler depend on %
variables an application
cannot fully control.
For example, the behavior of other applications executing at the same time,
high-priority tasks executed by the \os\ drivers,
or attempts to avoid priority inversion.

It is possible to capture all the allocator-related activity during the execution
of a critical phase for an entire process.
We find that this approach is fragile and noisy.
Some of the critical phases we measure last only tens or hundreds of milliseconds.
Executing even one unrelated callback thread in parallel with the critical phase
we measure can change the outcome of measurements by orders of magnitude.
We find it optimal to measure parts of the critical phases on a per-thread basis and later merge the results.

\subsubsection*{Custom memory managers}

Each \os\ provides a default user mode memory manager.
There are highly specialized applications and environments where performance is of the utmost importance (e.g., a database engine or a stock trading system)~\cite{evans_scalable_2011}.
Those applications benefit from custom memory allocators~\cite{berger,durner}.
We can use a custom memory allocator to increase the application's performance.
Configuration for a custom allocator can be tuned to optimize a specific application's usage patterns.
Custom allocators can use various techniques to intercept and replace the calls to the memory manager.
Those techniques may not be compatible with an interception of function calls to a default allocator we use.
We need to use the intercept methods specific to a custom allocator.

\section{Implementation related problems}
\label{sec:empimp}

When prototyping this methodology on application X,
we encountered the following categories of technical challenges:

\begin{itemize}
	\item \emph{Keeping track of allocations}. To keep track of relevant
	statistics, an application will need to store
	the memory usage patterns.
	We found that using data structures of a fixed size, or thread-local
	storage are functional solutions.
	If thread-local
	storage entries themselves are allocated lazily using
	\texttt{malloc()}, then explicit avoidance of recursion is also required.

	\item \emph{Avoiding performance impact}. To gather representative data,
	internal testing is not enough, and it is desirable to collect the
	metrics from the production environment. We observe that
	depending on a scenario, the number of interactions with the allocator
	may reach up to thousands or tens of thousands of events per second.

	\item \emph{Interactions with other intercept mechanisms}.
	Using multiple interceptors at the same time causes problems.
	Tools like {A}ddress{S}anitizer~\cite{asan_2012}, which engineers employ to detect
	memory corruptions, use their mechanisms
	to intercept memory allocation functions. Similar problems are
	associated with using profilers such as {X}code's {I}nstruments.
\end{itemize}

\section{Questions to research community}
\label{sec:questions}

Our findings have raised several questions which we seek answers to:

\begin{itemize}
	\item \emph{Relationship between memory allocator churn and other
	performance metrics}.
	Allocating memory is a process that causes the execution of \textsc{CPU} instructions.
	Depending on the \os, it may also cause noticeable delays in \textsc{I/O}
	operations in case of page faults.
	Our primary experience comes from i{OS}, where writable pages are never stored
	on the disk.
	Therefore, we have limited data about associating memory allocator churn
	with the cost of \textsc{I/O}.

	\item \emph{Determining \os\ and allocator-specific cost functions}.
	Based on our observations, the relative cost differs depending on \os,
	its version, compiler used, compiler options applied, or memory manager
	settings.
	We need more comprehensive testing and potentially benchmarks to evaluate
	how the weight of each of those functions depends on a specific environment.

	\item \emph{Replicating our work in different environments}.
	Our work is specific only to the environments where
	an increase in tens or hundreds of milliseconds is critical.
	We theorize that our proposed approach is also practical on products
	where the performance constraints are more relaxed.
\end{itemize}

\section{Threats to validity}
\label{sec:threats}

The threats associated with \emph{construct validity} are caused by
not interpreting or correctly measuring the theoretical constructs
discussed in the study.
We intercept only a specific set of functions related to memory
management.
Applications can always use lower-level functions directly and
bypass the runtime library exposing the memory management functionality.

One of the concerns for \emph{internal validity} is the interpretation of
results and if the conclusions we present can be really drawn from the
data available.
\emph{Because of confidentiality reasons, we are not able to
share either the specific performance regressions or their correlation
with other application-specific data with the public}.
We reach our conclusions based on
analyzing the data gathered from the production environment,
correlating it with other performance related metrics,
and discussions with performance engineering experts working on this topic.

Threats to \emph{external validity} are related to the application of
our findings in other contexts.
We have used this technique in the context of only one mobile \os,
a single application (albeit having hundreds of millions of users), and
a small set of programming languages ({C}, {C}++, and {O}bjective-{C}).
As a result, we cannot conclusively state that this approach works
for other environments.

\section{Conclusions and future work}
\label{sec:conclusions}

We present an approach agnostic of \textsc{CPU} execution
speed to understanding the evolution of
critical phases of software based on analyzing the memory
allocator churn.
The proposed technique enables comparison between execution intervals where
median duration in the production environment is from tens to hundreds of milliseconds.
Using memory allocator churn is a more deterministic
indicator of performance regressions than elapsed wall-clock time.
We use this technique for six months on an i{OS} version of application X.
Findings from the memory allocator churn help identify individual performance regressions and confirm (or disprove) the presence of problems uniformly impacting the entire application.

Our future work will focus on
\begin{enumerate*}[label=(\alph*),before=\unskip{ }, itemjoin={{, }}, itemjoin*={{, and }}]
    \item the toolset necessary for preemptively detecting
    the change in the allocator churn and notifying the engineers
    during the development phase

    \item determining the ways to rank the cost of functions
    based on their memory churn and provide engineers with means to
    assess the potential cost in the production environment.
\end{enumerate*}

\begin{acks}
    The authors are grateful to Gautam Nayak from Meta Platforms, Inc. for encouraging and supporting this work.
\end{acks}

\balance

\bibliographystyle{ACM-Reference-Format}
\bibliography{memory}

\end{document}